# Selling sex: what determines rates and popularity? An analysis of 11,500 online profiles

Alicia Mergenthaler[a], Taha Yasseri*[abc]

[a]Oxford Internet Institute, University of Oxford, Oxford, UK; [b]School of Sociology, University College Dublin, Dublin, Ireland; [c]Geary Institute for Public Policy, University College Dublin, Dublin, Ireland

*Corresponding Author: Taha Yasseri: taha.yasseri@ucd.ie

## Abstract

Sex work, or the exchange of sexual services for money or goods, is ubiquitous across eras and cultures. However, the practice of selling sex is often hidden due to stigma and the varying legal status of sex work. Online platforms that sex workers use to advertise services have become an increasingly important means of studying a market that is largely hidden. Although prior literature has primarily shed light on sex work from a public health or policy perspective (focusing largely on female sex workers), there are few studies that empirically research patterns of service provision in online sex work. This study investigated the determinants of pricing and popularity in the market for commercial sexual services online by using data from the largest UK network of online sexual services, a platform that is the industry-standard for sex workers. While the size of these influences varies across genders, nationality, age and the services provided are shown to be primary drivers of rates and popularity in sex work.

## Introduction

In this article, we analyse a dataset from AdultWork.com, a UK-based online platform to determine drivers behind pricing and popularity in the market for sex work. Our main research question was what are the determinants of price-setting and popularity in online commercial sex work? AdultWork was established in 2003 and is a platform where individuals pay to establish a profile online to sell sexual services. These services range from "cam services" (e.g. sexual services over live video), to erotic art, to in-person sexual services (AdultWork 2019). Each seller has a profile that provides information about the services they provide, details about themselves, the pricing of various services (rates) and the number of views on their profile. This data is used to analyse drivers of rates and popularity, while exploring the intersections of gender, nationality and service provision.

## Literature Review

Sex work, famously referred to by Rudyard Kipling in 1888 as the 'most ancient profession', is ubiquitous and poorly understood (Mattson 2015). It may be defined as the sale of sexual services for money or goods and exists in many forms (Adriaenssens et al. 2016). Data on sex work is notoriously unreliable and difficult to collect due to varying legality, stigma and the tendency for individuals to move in and out of sex work (Balfour and Allen 2014). The market for sex work is significant worldwide; the Office for National Statistics estimated the market for sex work in the UK at 5.3 billion GBP in 2009 (ONS 2014). Despite the largely hidden nature of sexual services, it is estimated that 85-90 percent of all sex workers are women, with an estimated 32,000 workers in London and approximately 72,800 in the UK as a whole (Balfour and Allen 2014; House of Commons 2016). Even in countries that allow the legal sale of sexual services between consenting adults, many sex workers remain secretive, as the profession is 'decriminalised but not legitimate' (UK Home Affairs Committee 2017). Moreover, estimating the number of individuals in sex work online is difficult in part because the number of profiles or adverts do not map cleanly to unique workers (Sanders et al. 2018). While terminology around selling sexual services is widely debated, the term 'sex worker' is used in this article to refer to individuals of all gender identities who exchange sexual services for money or goods.

The majority of older literature written on the topic of sex work has been small-scale, qualitative, or focused on specifically epidemiological applications (Rocha, Liljeros and Holme 2010). Recently, more research has been conducted on the experiences and market aspects of sex work due to the accessibility of online sex work platforms. This emerging body of literature on this topic has not thoroughly examined important intersections between marketplace success, male gender, nationality, age and the services that sex workers offer (Jones 2015). This study seeks to enrich existing research examining the important interactions between sex worker attributes, price-setting, and popularity.

Prior literature reflects many heterogeneous processes that can lead to individuals being involved in sex work: financial benefit, homelessness and drug addiction, family breakdown, and trafficking (Balfour and Allen 2014). While some studies report that a majority of sex workers were content with their working conditions and control over their work (Mai 2009), others report workers (especially street workers and victims of trafficking), being unhappy with their conditions and wanting to change their situation (Scambler 2007) .

Bettio, Della Giusta and Di Tommaso (2017) posit that the agency and stigma of sex workers exists in a continuum across the market. Agency is defined as the ability to 'influence the terms of exchange' (Bettio, Della Giusta and Di Tommaso 2017), such as having control over working hours, working conditions and choice of clients. Workers at the higher end of the market experience (escorts) may experience greater agency and less stigma than their lower-end counterparts (street workers) (Bettio, Della Giusta and Di Tommaso 2017). Furthermore, reasons for entering sex work, completeness of information, and degree of risk associated with the sex trade varies dramatically between workers who work on the street and those who work indoors (Bettio, Della Giusta and Di Tommaso 2017). As Platt et al. (2011) found in a cross-sectional study of sex workers in London, migrant workers in the UK from Eastern Europe and the former Soviet Union face increased risks of violence and sexually transmitted infections (STIs) in comparison to their UK counterparts. There are additional differences in motivation between female, male and transgender workers (Balfour and Allen 2014). It has been reported that a detailed analysis of male and trans-identifying sex workers is lacking in studies of sex work (Jones 2015). Prior studies have posited that the male sex market is largely linked to the commercial gay sex scene and that there are very few women who buy sexual services (Balfour and Allen 2014). However, this position has been nuanced by the analysis of some specialised heterosexual-focused platforms which contain a larger population of straight male-identifying sellers (Sanders et al. 2018).

The increase in popularity of online platforms has transformed sex work by providing sex workers with more agency over their work, a wider variety of opportunities (such as purely online services like 'camming') and greater profit potential (Cunningham et al. 2018). Furthermore, the emergence of the Internet has influenced the market for sexual services by facilitating for the exchange of detailed information between buyers and sellers and allowing for reviews and other signals of seller reputation (Cunningham and Kendall 2016). By analysing Internet platforms for selling sex, there is potential for understanding the experiences of sex workers and the larger escort industry. Disentangling the drivers of rates and popularity of sex workers can help illuminate the value placed on sexual services, and the experiences of sex sellers as a whole.

The analysis of sexual marketplaces is especially illustrative of some aspects of sexual preferences, as these selections happen in private in contrast to dating preferences, in which there are significant status effects to public mate selection (Cunningham and Kendall 2016). Platforms that mediate the exchange of sexual services may allow consumers to find escorts and services that better match their preferences (Holt, Blevins and Fitzgerald 2016). These lowered search costs have been especially effective in matching consumers with high-end escort services and "girlfriend experiences" (Cameron 2016). The ability to target more specific and “elite” clients through advertising can help sex workers maximise their profits (Cunningham et al. 2018). In Holt, Blevins and Fitzgerald (2016), the authors also posit that information on the Internet allows buyers to select for racial preferences and specific kinds of sexual services.

Increasingly, research has been done to understand the pricing of sexual service and possible influences of popularity in the market by using online advertisements and sex work platforms (DeAngelo et al. 2019; Cunningham et al. 2018). In DeAngelo et al. (2019), the authors suggest that sex work prices are determined largely on the supply side and designed to compensate for risk. Sellers charge a high premium for traveling to a location of the buyer’s choosing, and sellers set rates that correlate strongly to local violent crime rates (DeAngelo et al. 2019). Qualitative studies also suggest that the type of service and the setting of service influence rate-setting. The provision of unprotected sex in particular may lead to differences in pricing. Elmes et al. (2014) suggest that the rate for unprotected sex is on average higher than protected sex. In their study, which examined rate differences of unprotected sex for sex workers in Zimbabwe, the authors observed that clients paid on average 42.9 percent more for unprotected sex (Elmes et al. 2014).

In selecting worker attributes that could influence rates and popularity, we considered the existing literature on factors influencing sexual preferences. These preferences are highly dimorphic across genders; in one study of online dating behaviours, researchers could predict gender from mate preference with 91 percent accuracy (Conroy-Beam et al. 2015). Features like age, body type and race can influence sexual selection. Male preferences for younger female sexual partners, for instance, are reflected robustly across a large body of literature. On average, men prefer a lower age for their mates than women do, and this finding is consistent across a study of 27 cultures (Grøntvedt and Kennair 2013; Buss and Schmitt 2018).

Body type is also an important feature in mate-selection. Research on adults has indicated that individuals have preferences for certain waist-to-hip ratios and specific builds in potential mates. For instance, the preference for women to have relatively low BMI and a low waist-to-hip ratio has been empirically proven by many independent studies (Buss and Schmitt 2018). Other studies suggest that men, especially bisexual and gay men, experiencing stigma about having high weight in dating and other contexts (Austen, Greenaway and Griffiths 2020). This is further reflected by the bias against men believed to be overweight in gay and other men who have sex with men sexual marketplaces (Logan 2010; Goldenberg, Vansia and Stephenson 2016) .

There is significant literature on the influence of race and ethnicity in dating preferences. Research indicates that same-race preferences exist in dating, and that gender, age, background and attractiveness can influence the strength of this preference (Fisman et al. 2008; Lundquist and Lin 2015). In Fisman et al. (2008), the authors found that women had stronger same-race preferences than men, and older subjects had weaker same-race preferences than younger subjects. The magnitude of this preference may vary across individuals of different sexual orientations and races, with one study showing that white gay men and lesbians displayed the greatest openness around dating outside their race (Lundquist and Lin 2015).

Nationality may also influence selection and rates in the marketplace through another mechanism. Cunningham et al. (2018) found that migrant workers are at higher risk of xenophobic and racist comments on customer review forums. In light of Brexit in the UK, it is possible that increased xenophobia would bias buyers against sex workers reporting non-UK nationality; these workers already experience a greater incidence of violence than their British counterparts (Platt et al. 2011). In addition, the UK has a large number of migrant sex workers (50 percent of all female sex workers), some of whom entered the UK against their will (5 percent), and 25 percent of whom come from Eastern Europe and the former Soviet Union (Platt 2009). Finally, migrants may struggle with access to the necessary technology to facilitate their online sex work or experience greater language barriers in setting up their profile and businesses (Sanders et al. 2018). A greater degree of vulnerability and economic insecurity among migrant workers may influence their rate-setting.

## Data and Methods

### *Data Collection*

Each profile on the AdultWork.com contains information about the worker and the services they provide. While variable across providers, a profile typically includes personal information about the seller (age, gender, orientation, location, race, nationality, body statistics), a free-text component where workers can write more about themselves, a photograph, and a link to contact the seller. In order to collect the data from AdultWork, we scraped raw HTML files of seller profiles from the site and used an Extensible Markup Language (XML) parser in Python called ElementTree to extract features from the profiles. We collected all available fields from the profiles, with the exception of telephone numbers and pictures.

*Ethics*

The collection of data analysed in this study is in line with rules laid out by the Intellectual Property Office of the UK and may be analysed because it is freely available and public, requiring no log-in (Van Hoorebeek 2014) and individuals have willingly posted this information publicly. Participants are protected because the findings are only presented in aggregated format, no usernames are collected, and all data is held in an encrypted form. The research protocol has been reviewed and approved by the Central University Research Ethics Committee (CUREC) and Humanities Inter-Divisional Research Ethics Committee (IDREC) at Oxford University (SSH_OII_CIA_19_016).

### *Data Cleaning*

Figure 1 shows the process gone through to select profiles for the study. We started with 33,002 profiles and ended with 11,489 profiles after applying multiple filters. Sellers who did not offer any in-person services on their profiles were excluded from the analysis. A qualitative inspection of these profiles revealed many of these sellers either had deactivated accounts, were the purveyors of erotic art, or provided virtual services (e.g. "cam", or video services). Sellers that provided an hourly rate on their profile were included in the study, and the remaining profiles were excluded. We also excluded workers posting implausible information on their profiles such as reporting an age over 100. Finally, we excluded the small number of sellers not reporting a gender, who were part of a couple, or who were outside of the UK.

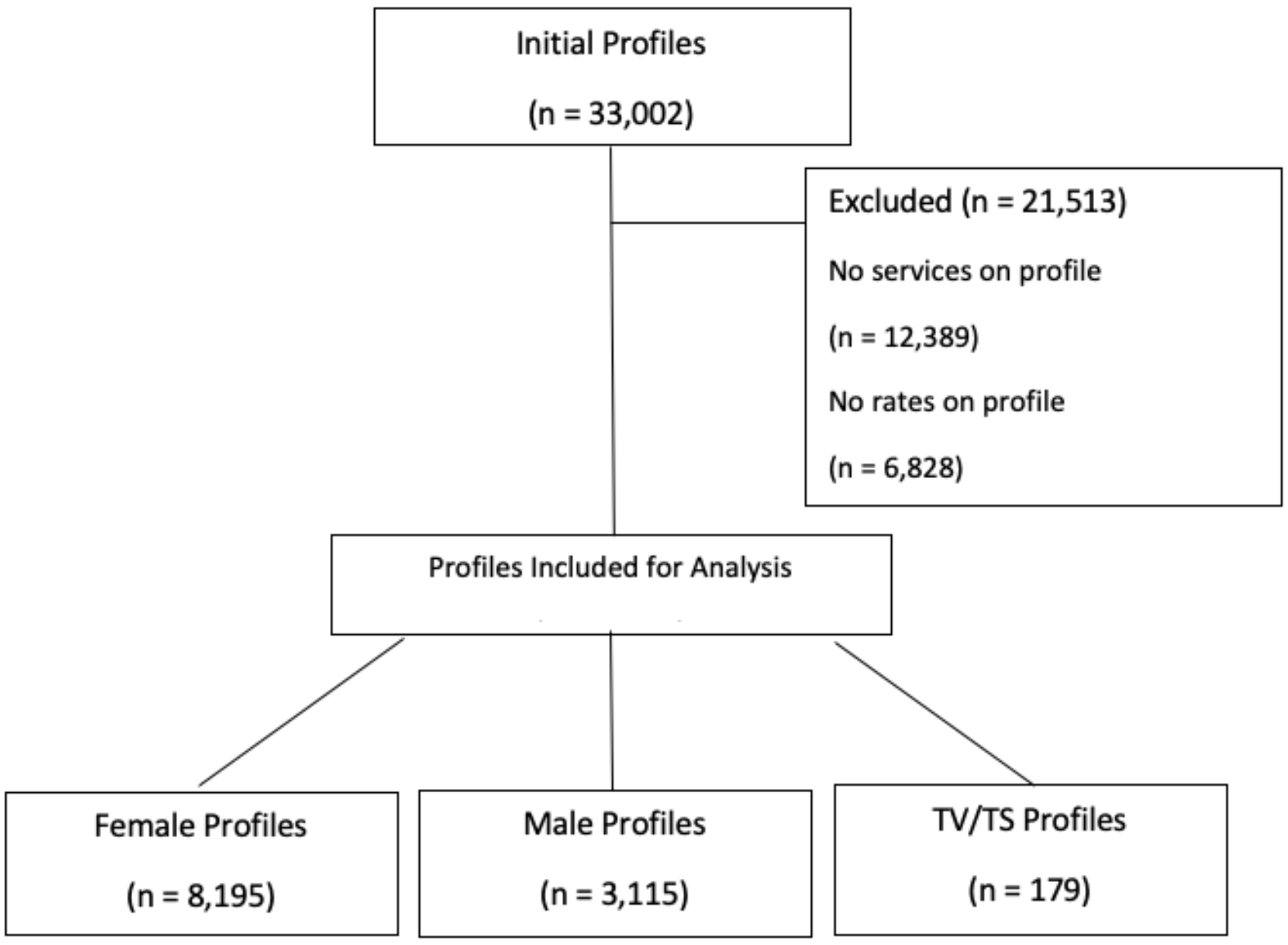

**Figure 1.** Profile selection procedure.

### *Variables*

We used profile views, a variable that is populated automatically by the website, as a proxy for worker popularity. We normalised this metric and reported views per day to control for profile age. Rates are determined by the provider and are a representation of self-perceived attractiveness in the market. Each seller provided a maximum of 16 rates for in-calls and out-calls for different durations of service. The durations of service were "15 minutes", "30 minutes", "1 Hour", "1.5 Hours", "2 Hours", "3 Hours", "4 Hours", and "Overnight". "In-call" refers to the scenario in which the client goes to the sex worker’s place of residence or work. Conversely, "out-call" indicates a situation in which the sex worker travels to a specified location outside their place of work (Griffith et al. 2016).

In addition to rates and profile views as dependent variables, we selected nationality, gender, time active on the website, age, ethnicity, region, orientation, dress size and services offered as primary independent variables. A summary of the variables is shown in Table 1.

**Table 1.** Summary of variables.

| Variables | Definition | Number of Possible Values |
|---|---|---|
| 1 Hour IC | 1 Hour In-Call Rate | Continuous |
| Views | Profile Views Per Day | Continuous |
| Gender | Male or Female | 2 |
| Nationality | Grouped Nationality | 9 |
| Ethnicity | White, Non-White | 2 |
| Age | Current Age of Worker | Continuous |
| Region | London/Southeast and Other UK | 2 |
| Orientation | Straight, Bisexual, Gay, Not Specified | 4 |
| Unprotected | Offering "Bareback" or "Unprotected Sex" Service (0 or 1) | 2 |
| Dress Size | Grouped Ordinal Sizes | 3 |

### *Service Provision*

Providing a description of services provided is compulsory for sellers on the platform. 90 different services are offered across the platform. The most common services on the platform include practices that appear on more than 10,000 of the profiles (e.g. massage and oral). The least common services provided appeared on as few as 8 profiles (e.g. hardsports).

*Clustering services*

Building from the hypothesis that some services commonly appear on profiles together, we developed a method to group services into categories. To investigate this hypothesis, we constructed a network connecting services that commonly appeared together, and grouped these services using a community detection algorithm. More details about this methodology, along with the network diagram of services, exists in online Supplementary Information Text S1 and Figure S1. Ultimately, this method yielded four different communities, or groupings of services. We refer to these as categories A-E. Category A contained services that exist in most profiles, e.g. massage and oral sex ("common services"). B contained more social and collective acts. These included services such as swinging, parties and moresomes ("group/ public services"). C contained mostly BDSM-related services. D encompassed increasingly more niche services, e.g. enema, rimming, snowballing ("niche services"). Finally, E constituted even rarer services that did not belong to any one community, but highly overlap with each other (unprotected sex and period play). Notable in this category was the explicit offering of unprotected sex. Different services and the number of workers that offer them can be seen in Figure 2.

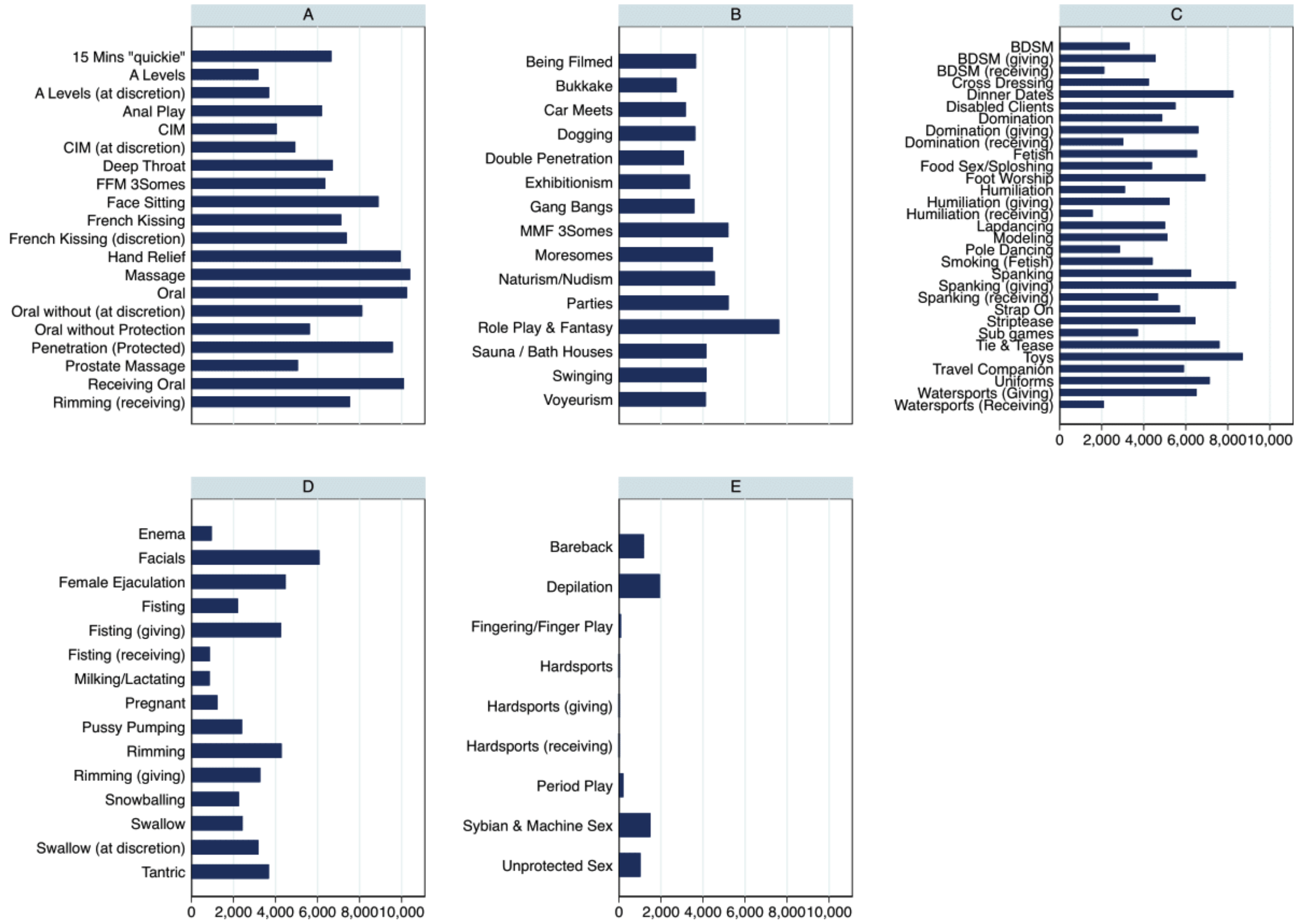

**Figure 2.** Number of workers who offer services by category.

## Results

### *Exploratory Analysis*

*Gender*

Individuals on AdultWork were divided into the categories: female, male, and TV/TS. The gender breakdown was 71.3% female, 27.1% male, and 1.6% TV/TS. The distribution of rates and views varied dramatically between men and women. Trans-identifying sellers were left out of our statistical analysis due to the small group size, however, it is notable that while estimates of the transgender population in the UK range from 200-400,000 individuals, 0.3-0.6% of the population (ONS 2017), the fraction of trans-identifying individuals on AdultWork was 1.5 percent of the total population.

*Views, Rates and time active on platform*

Sellers had profiles of different ages on the platform. We calculated the amount of time active spent on the platform by the amount of time between last log-in and the date of profile creation. Figure 3(a) in online Supplementary Information shows the distribution of active years on the platform. Most of the sellers on the platform had approximately 2-3 years of activity. The median length of active time was 3.7 years. The maximum length of activity was 16 years. As AdultWork was founded in 2003, this indicates that a small number of individuals have been active on the site since near the time of its founding.

The number of profile views is a significant metric for sellers at the "top of the sales funnel". Buyers need to view a profile in order to know their availability and contact them. The distribution of views per day can be seen in online Figure 3(b). Views roughly followed a log-normal distribution with two modes for male and female profiles and with a lower limit for men. The median number of views per day for a female and male seller was 296.4 and 2.6 views per day.

Hourly rates on the platform are determined by the provider’s incentive to stay competitive in a market where the rates of other providers are visible. Sellers may also set rates to signal quality or vertically segment themselves in the market (Cunningham and Kendall 2016). Online Figure 3(c) shows the distribution of the 1 hour in-call rate in the male and female sub-populations in GBP. Note the logarithmic scale, which suggests a fat-tail for the distribution with a very wide range of rates within each sub-population. The median price of a 1 hour in-call was 120 and 70 GBP for female and male workers. This finding is notable, especially in the light of UK statistics reporting that the average price of an encounter with a sex worker is 78 GBP (UK Home Affairs Committee 2017).

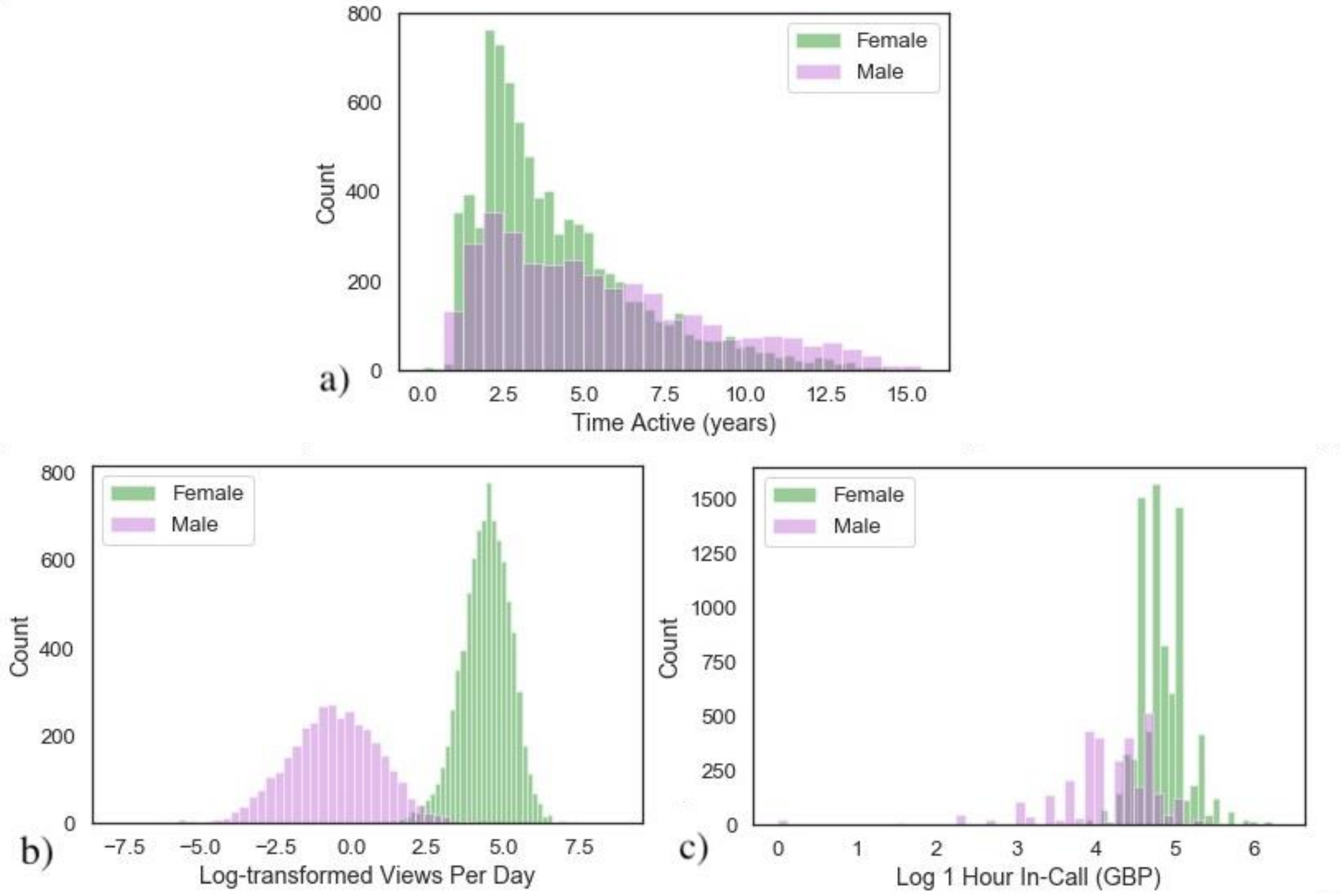

**Figure 3.** Distributions of a) the time active on platform, b) daily profile views, and c) hourly in-call rate. All variables are logarithmically transformed.

A one hour in-call was the most popular service provided by sellers. The least frequent duration was 15 minutes, indicating a possible minimum time threshold in which it is worth it for a seller to provide services. Moreover, workers were more likely to offer durations shorter than an hour for in-calls than for out-calls. This could indicate greater opportunity cost on behalf of the seller when they have to travel to meet a client. Similarly, lower hourly rates were observed with longer services. In-call and out-call rates are linearly related (See Figure S2 in online Supplementary Information). However, out-calls were consistently more expensive than in-calls (1.26 times larger on average) corroborated by past research on the advertisements of US sex workers (Griffith et al. 2016; DeAngelo et al. 2019).

*Nationality, Age, and Dress size*

Nationality is not a compulsory data field; only 14% of men and 64% of women provide their nationality on their profile. Among those who provided a nationality, 93% of men were "British". Women had a much more diverse distribution of nationalities, with 65% "British" followed by "Romanian", "Hungarian", "Brazilian", "Czech", "Polish", "Thai" and "Spanish". We grouped nationalities into 9 larger regions: Britain, Eastern Europe, Western Europe, Asia, Central/South America, Africa, North America, Oceania, and the Middle East (See Table S1 in online Supplementary Information). The counts of female workers by nationality can be seen in Figure 4(a).

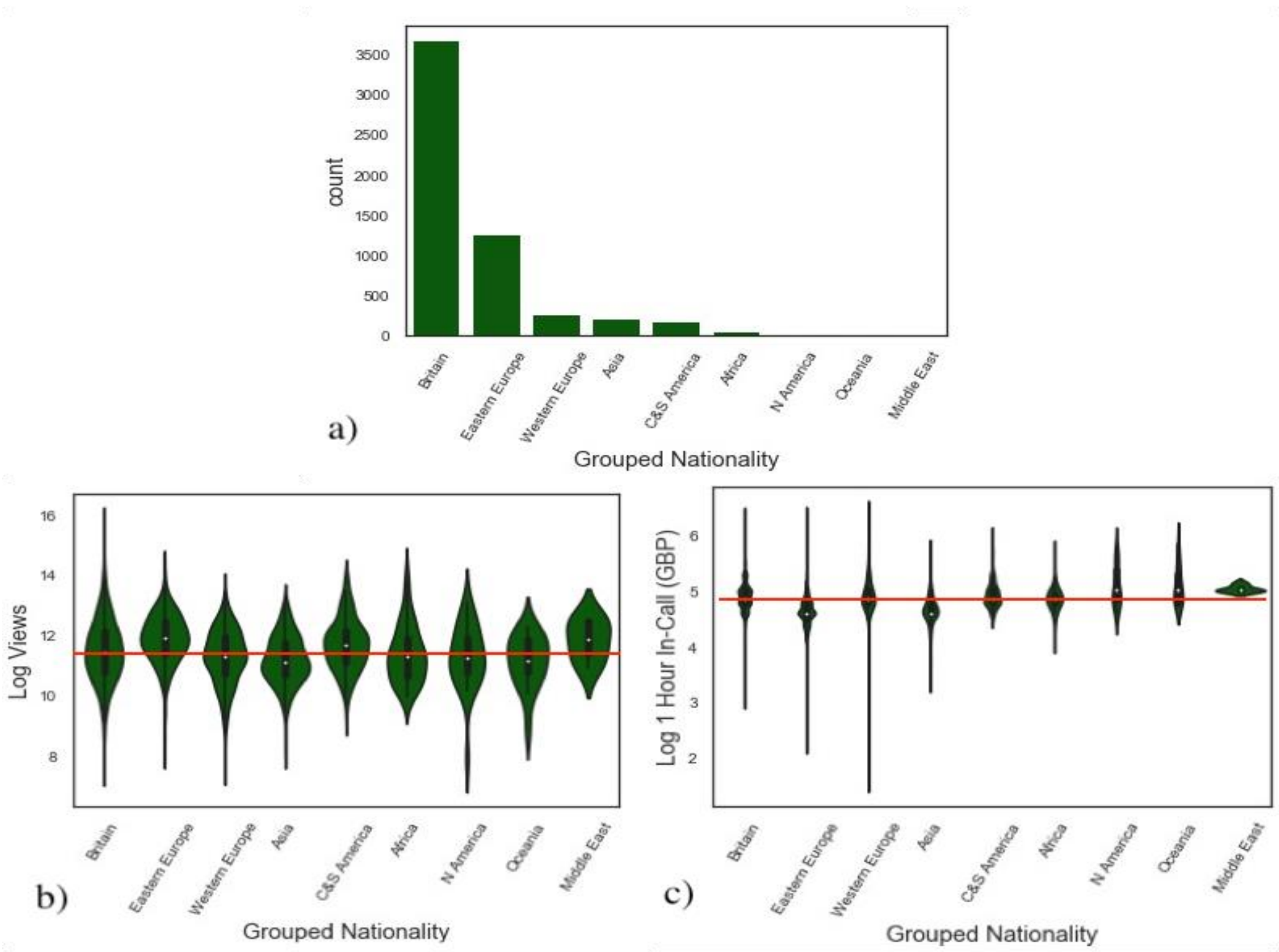

**Figure 4.** a) Grouped nationality counts, b) views, and c) rates.

The distributions of views and rates are detailed in Figures 4(b-c) for different nationality groups. Female Eastern European sellers garnered slightly higher median number of views than their British counterparts (408 per day vs. 256 a day), indicating that they were receiving equal or even more browsing attention than British sellers. More notably, workers of Eastern European nationality had significantly lower average hourly rate than their British counterparts. The median 1 Hour In-Call price for female British sex workers was 130 GBP, whereas female Eastern European sellers had a median 1 Hour In-Call rate of 100 GPB (30% less). As nationality is impossible to verify, there could be incentives for sellers to omit or misreport their nationality. This is especially plausible given the large price difference between providers.

Age is a required field on the platform. The platform buckets sellers into eight ordinal categories on their search bar. The median age for men and women is 35 and 32 years. The dependency of views and rate on age is shown in Figure S3 of the online Supplementary Information. For women, there is a penalty in rates and views for moving up an age category. This relationship is initially less clear for male workers. We also calculated the age at which the sellers joined the platform by subtracting the profile age from the current age of the sellers. The distribution peaked around 22 for female and 25 for male sex workers (See Figure 5).

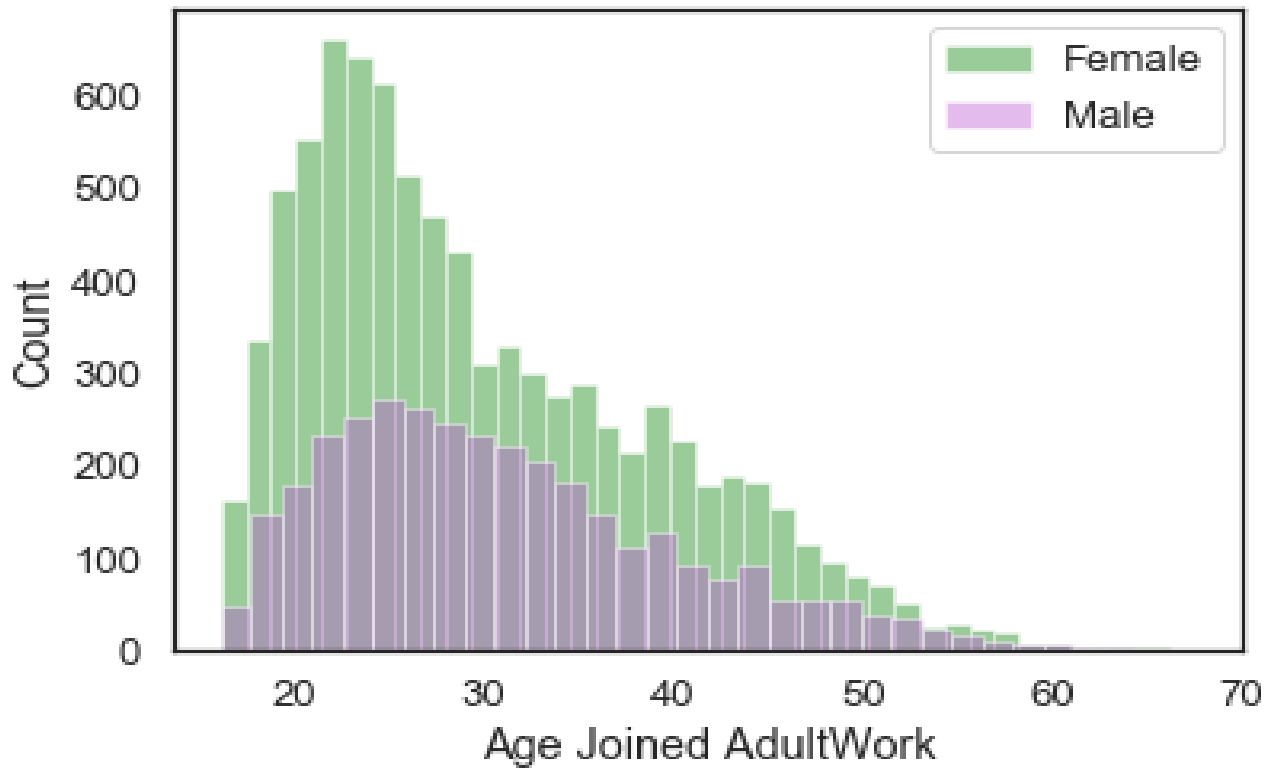

**Figure 5.** Age of joining the platform by gender.

We used dress size as a proxy for body type for women, as the website does not offer more specific fields for BMI or weight. Dress size is optionally reported on the website. 73% of women report a dress size, whereas less than 1% of men did so. Hence, we only consider female workers' dress size. The most frequent dress size reported on the platform was a UK size 10 (see Figure S4; for comparison, the average dress size of 16-24 year-old women in the UK in 2016 was a 14(*BBC*, March 3, 2016). As shown in Figure S4(b-c) views and rates decrease as dress size increases. Table S2 in the online Supplementary Information contains an international dress conversion chart for reference.

*Region, Orientation, Ethnicity*

Workers who reported their region (56% of all profiles) were concentrated in London (20%) and the Southeast (35%). Exploratory analysis revealed that the largest difference in 1 hour in-call rates occurred between London and the Southeast, compared to other regions of the UK. Because of this, we divided the regions into two groups: London and the Southeast (LS) and Other UK. Sex workers in LS had a higher hourly in-call than their counterparts in other parts of the UK. The profile views per day was independent of region. Female workers in LS and Other had a median views per day of 305 and 304 respectively, and a median rate of 130 and 120 GBP. Male sex workers had a median views per day of 3 and 2.4 and a median rate of 80 and 70 GBP in LS and Other UK respectively; see more details in the online Supplementary Material - Figure S5.

The ethnicity of 50 percent of sellers on the platform is self-reported as white with 21.2 percent missing entries and roughly 21.8 percent reporting as non-white (see Figure S6 for more detail). Ethnicity is forced into categories by the platform. Non-white sellers report as majority mixed, Black or Asian. The large number of missing entries could be in part due to the platform affordances. It was evident that at least for female workers being "non-white" brought fewer views and a lower hourly rate.

The orientation of 57.3% of sellers was self-identified as bi-sexual or bi-curious, with 41.9% identifying as straight, and less than 1 percent identifying as gay. The large number of bi-identifying individuals in the data could be advantageous from the perspective of sellers attempting to reach as wide of an audience of prospective buyers as possible. Sellers advertising as bisexual or bi-curious received a modest boost to popularity and rates on average, in comparison to other orientations (see the online Supplementary Material - Figure S7 for more details).

### *Regression Results*

We ran regression analyses to model views per day and hourly rate of workers by gender sub-population. The reader should be reminded that these two variables are proxies for initial attractiveness to buyers and self-assessed price by sellers. Considering the shape of distribution of these variables, we used logarithmically transformed values of them in our regression analysis.

*Female Workers*

In analysing the female sellers on the platform, the baseline comparison groups for binary variables were British nationality, white ethnicity, dress size 4-8, and straight orientation (Table S3 in Supplementary Information). Notable among the regression results were the negative effects of age on hourly rate. The former has been documented in previous literature for female sex workers (Dunn 2018). Ethnicity played an important negative role on viewership, whereas it had no significant effect on the hourly rate. This finding is in agreement with prior literature that finds that buyers demonstrate racial bias in selecting sex workers (Holt, Blevins and Fitzgerald 2016; Robinson 2015).

Female workers were heavily penalised on their dress size. Notably, the negative effect of this on their chosen hourly rates was larger than on their viewership. Nationality, however, plays the most important role in determining the hourly rate. For the majority of non-British nationalities, rates were lower compared to workers of British nationality. The largest decrease in hourly rate was imposed on workers from the East European and Asian countries followed by South and Central Americans. A missing nationality was associated with a decrease in both dependent variables.
After controlling for the total number of services provided, female workers offering category C ("BDSM services"), category B (collective or external-facing services such as dinner dates and threesomes) and category D ("niche services") charged higher hourly rates. Providing services in LS region was associated with higher rates on average than workers providing services in Other UK. Finally, identifying as bisexual (versus a baseline of straight) was associated with a modest boost to hourly rate on average.

*Male Workers*
We ran a similar regression model for male workers without the Dress Size variable (Table S4). We observed less explanatory power in the selected variables for male workers. This suggests that the role of demographic and physical attributes in popularity and hourly rate of sellers is less important when it comes to male workers. This finding is in line with other research on male sex workers, where popularity is generally less readily predictable based on physical features for male workers than for female workers (Logan 2010). Nevertheless, similar to female sellers, older male sellers charged lower hourly in-call rates. Dissimilar to female workers, ethnicity played no role in rate-setting in the case of male workers.
The most notable nationality effect came in the case of Central/South American workers who had significantly lower hourly rates than British workers. However, one should remember that the male population in the sample is predominantly British and reported results might not be easily generalised to a more balanced population. As for the services provided, male workers that provided more "A", or common services, tended to charge a lower hourly rate. The regional trend of offering lower hourly rates in Other UK regions continued with male sellers.
It is worth noting that the selection advantage of identifying as bisexual for female workers disappears in the male population, of which 62% identified as straight.

*Unprotected Sex*
Our dataset presents a unique opportunity to analyse the provision of unprotected sex in the marketplace. Offering unprotected sex has ramifications for pricing of sexual services (Cunningham and Kendall 2016; Elmes et al. 2014). Consistent with the findings of this literature, we find that providing unprotected sex is strongly associated with offering lower hourly rates in our dataset. We focus on the provision of unprotected sex because it has wider safety and epidemiological applications for workers. 1,304 providers offer either "Bareback" or "Unprotected Sex" or both, constituting 11.35% of the sample.
While some studies have predicted a higher premium for unprotected sex (Elmes et al. 2014; DeAngelo et al. 2019), potentially to compensate for the additional risk undertaken by the seller, others indicate a lower price for sellers offering unprotected sex (Cunningham and Kendall 2016). We found that sellers that offered unprotected sex services tended to offer less expensive 1 hour in-call rates (65 and 110 GBP for men and women) than their counterparts who did not offer these services (70 and 120 GBP for men and women) on average ($p < .05$). This corroborates the findings in Cunningham and Kendall (2016), who found lower rates on average for unprotected sex in a study of a US-based platform, the Erotic Review.
We used logistic regression to analyse the likelihood of providing bareback or unprotected sex services (see the online Supplementary Material -Table S5). The greatest predictor of unprotected sex is identifying as male. White providers, those from Eastern Europe, and younger sellers were also more likely to provide bareback or unprotected sex.

## Discussion

In this study, we found that nationality, age, services offered, ethnicity, region and sexual orientation influenced rates and popularity in the market for commercial sexual services. The influences of these variables are complex and differ dramatically between different the sexes. While all of the variables in the study add explanatory power to understanding the market for sex work, there is much more variability in the data that is left to explain.

The unique drivers of rates found in this research can help understand the lived experience of sex workers. Building on the work of Bettio, Della Giusta and Di Tommaso (2017), sex workers who offer lower rates are more likely to experience stigma and less agency as being part of the 'lower-end' sexual services market. In particular, among cisgender women in the UK context, Eastern European nationality is strongly associated with lower rates. This finding, along with the result that reporting Eastern European nationality is significantly associated with a higher probability of offering unprotected sex, empirically supports the findings of Platt et al. (2011), a small interview-based study highlights the health risks and stigma associated with being an Eastern European migrant sex worker in the UK.

The findings shed light on the relatively high prevalence of male sex workers, who have unique drivers of pricing and popularity. Understanding the scope of male sex workers is especially important for the allocation of future resources for sex worker safety and educated policy creation around these populations. While prior literature indicates that 85 to 90 percent of all sellers are women, this finding reveals how research consistently underestimates the importance and number of men in sex work (UK Home Affairs Committee 2017).

## Limitations

Our results suggest that there are additional variables not available in the dataset that could explain male price setting and popularity in the market. The existing model that includes sex, nationality, age and location explains only a modest amount of variability in the rates of male sex workers. These findings are consistent with Logan (2010), which finds that the patterns of rate-setting are markedly different between male and female sex workers. Moreover, Logan (2010) finds that personal physical characteristics are largely unrelated to price-setting other than those having to do with body build (e.g. muscular men have higher rates and overweight or thin men have lower rates). The findings in Goldenberg, Vansia and Stephenson (2016) also suggest that there is bias against "fat" or "ugly" men in Craigslist ads within communities of men who have sex with men (MSM). Sellers identifying as tops also demand higher rates while bottoms demand lower rates (Logan 2010). These findings suggest that having additional information surrounding sexual roles could help explain variation in male pricing. Finally, variables explicitly involving STI status (especially HIV status) and drug use were not available in our dataset. Additional information about disease and drug status could be useful for understanding pricing in future work (Goldenberg, Vansia and Stephenson 2016).

Other than website-specific features, the information provided by users is self-reported. Depending on the information and the perceived desirability of the trait, there may be systemic under or over reporting of features on the platform (i.e. if emphasising one trait is advantageous, sellers may have incentive to highlight that trait). It is worth noting, however, that being untruthful about traits like age that could potentially be detected by a buyer could hurt the seller in ratings, future transactions, and potentially subject them to danger (Holt, Blevins and Fitzgerald 2016). External validity is also a concern in the discussion of the market for sex workers. The individuals represented in this data do not necessarily represent the overall population for sex work, and they represent a selected subset of individuals who participate in online sex work environment.

## Future Research

Findings from this research invite further investigation into the nature of sex work in the UK. More work could be done to understand the clientele of the website given their preferences in the marketplace (i.e. the preference for lower ages reflecting mostly male customers). Finally, while ratings exist on AdultWork, a very small percentage of customers leave them and the figure is highly biased towards positive ratings. More work around reputation mechanisms and networked screening function in the market for commercial sex work could illuminate how customers and workers match with each other. Customer review forums are especially rich sources of information that could shed light on the well-being of sex workers. These forums serve as a “double-edged sword” (Cunningham et al. 2018), as they can substantially help sex workers with positive marketing or cause damage through online abuse, bad reviews, or the leaking of private information.

For future research, additional variables from AdultWork and similar websites could be used to build on our findings. Researchers can draw from the large amount of unstructured data in each profile description and the frequently asked questions on each page. Potential variables of interest could include the number of images on the profile or the length of the profile page description. Both of these variables could be indicative of both seller engagement and the amount of information available to consumers about each worker. Furthermore, gauging the similarity of profiles across the website in content and style could yield relevant information about who is creating seller profiles. If profiles show a high level of similarity, this could suggest that they had the same creator, or were created by a professional. Control over a profile could indicate organisation of sex workers (which is illegal in the UK), and also could be relevant given our finding that migrants and other vulnerable groups are at risk of exploitation on commercial sex work platforms.

## Conclusion

Our study seeks to investigate the factors that drives pricing and popularity in the market for commercial sexual services, a notoriously difficult topic to research. Several predictable variables influenced pricing and popularity, such as the age and location of the sellers. However, using the unique affordances of the AdultWork online platform, we found that services offered by sellers (including offerings like unprotected sex), as well as nationality, exert influence over pricing and popularity in the marketplace. Ultimately, selection based on services offered demonstrates the ability of buyers to exploit the versatility of options online with lower search costs (Cameron 2016).

## Acknowledgement

We thank Ieke de Vries for the valuable comments on the manuscript.

## Funding

TY was partially supported by the EPSRC grant no. EP/N510129/1.

## Data availability

All the data used in this work is publicly available via Harvard Dataverse:
https://doi.org/10.7910/DVN/OMCDPI

# Online Supplementary Information

## Selling sex: what determines rates and popularity? An analysis of 11,500 online profiles

Alicia Mergenthaler[a] and Taha Yasseri*[a,b,c]

[a]Oxford Internet Institute, University of Oxford, Oxford, UK; [b]School of Sociology, University College Dublin, Dublin, Ireland; [c]Geary Institute for Public Policy, University College Dublin, Dublin, Ireland

*Corresponding Author: Taha Yasseri Email: taha.yasseri@ucd.ie

### Text S1.

In order to compute service categories, we began with a matrix A of the dimensions n by m where n = 11, 489 is every worker in the dataset and m = 90 is the number of services offered on the platform. The vector of services, $a_{ij}$, takes on values of 0 or 1 depending whether or not each service appears on a worker's profile. Subsequently, we used the following expression to compute the cosine similarity between service column vector $\boldsymbol{a_i}$ and $\boldsymbol{a_j}$ as

$$\cos(\mathbf{a_i}, \mathbf{a_j}) = \frac{\mathbf{a_i} \cdot \mathbf{a_j}}{\|\mathbf{a_i}\| \|\mathbf{a_j}\|} = \frac{\sum_{k=1}^{n} a_{ik} a_{jk}}{\sqrt{\sum_{k=1}^{n} a_{ik}^2 \cdot \sum_{k=1}^{n} a_{jk}^2}}$$

Higher cosine similarity between services suggests that these services more frequently appear on profiles together controlling for their own frequency of appearance. Using the cosine similarity, we created a network with nodes consisting of sexual services and weighted edges consisting of the cosine similarity between services. We used the Louvain community detection method (Blondel, Guillaume, Lambiotte, and Lefebvre 2008), to find clusters of the services. Figure S1 contains a visualization of the network of services, with different colours representing different communities.

To understand the degree to which workers provide services from each category, we compared each worker's vector of services to each community's basis vector. As the vectors are orthogonal and no one service is present in more than one category, this method created five different scores for each worker that quantifies the similarity between the set of services they provide and the services in each of the categories A–E.

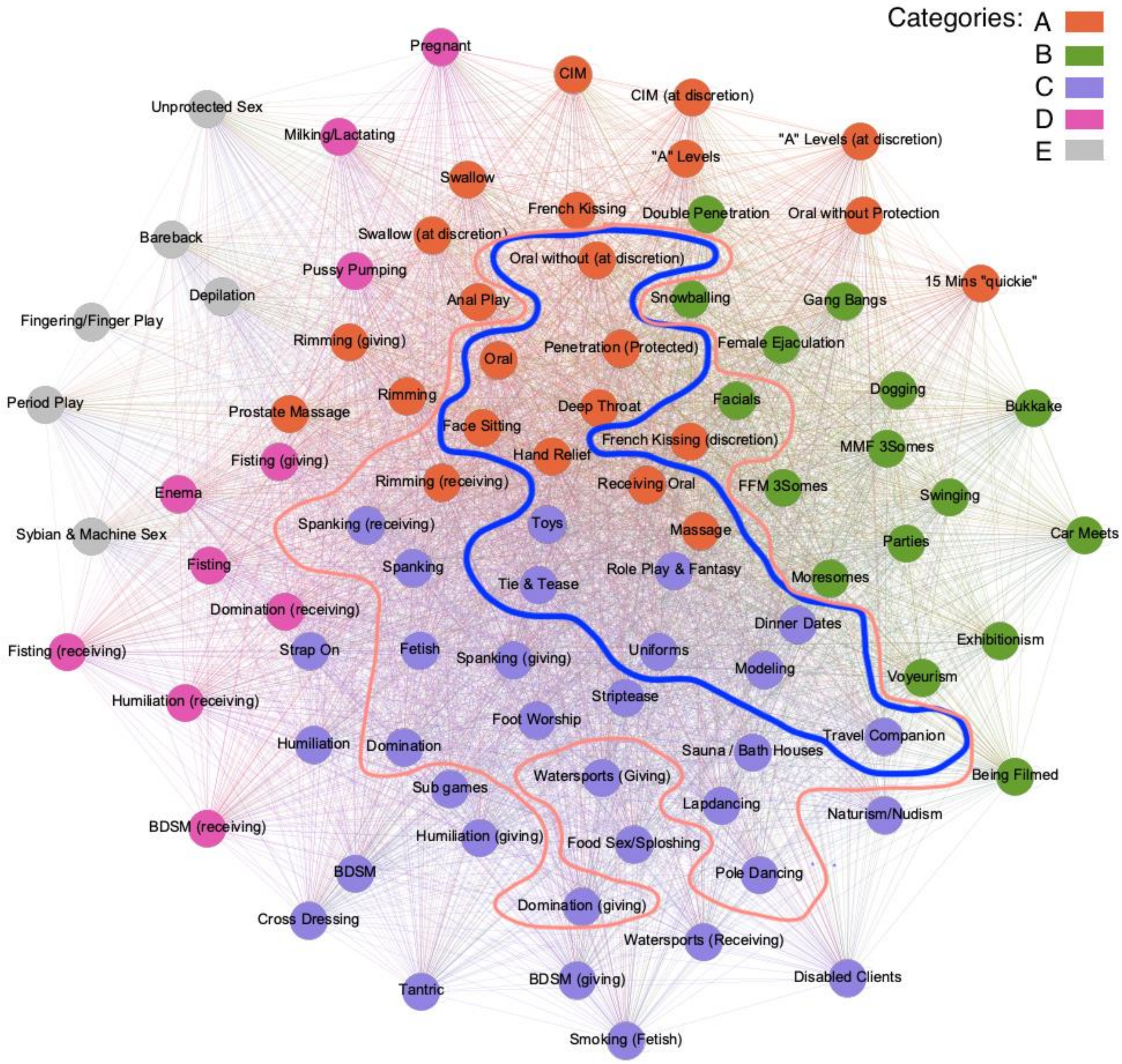

**Figure S1. Network of services offered.** The connections represent the co-occurrence of the pair of services on the same profiles, controlled for the popularity of each service (see Text S1). The blue and pink lines denote the core 15% and 30% of the services respectively, identified by filtering out the connections with smallest weights.

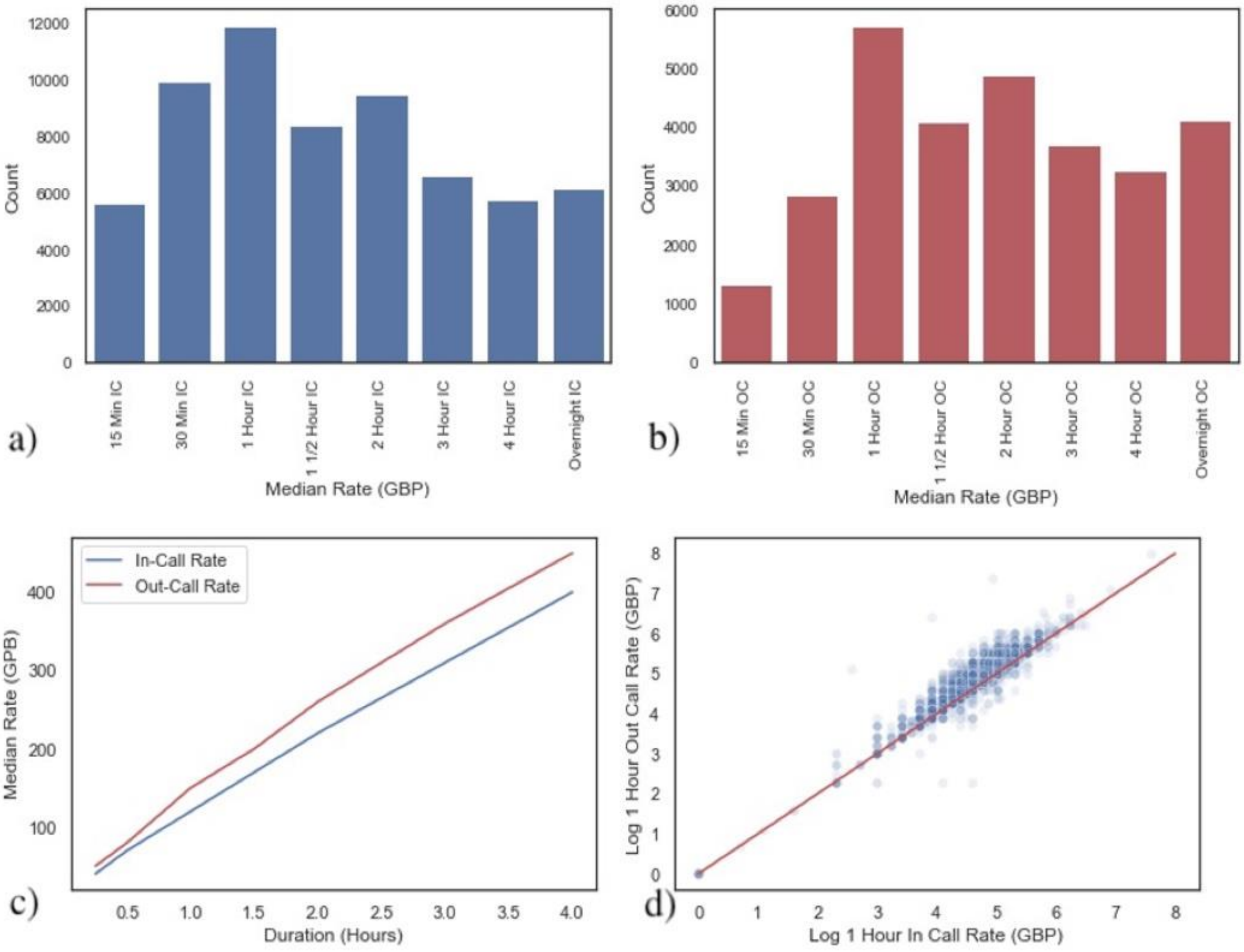

**Figure S2. Count of profiles with a set value rate for different durations of service.** a) in-calls and b) out-calls. c) Median rate over duration of service. d) Comparison of in-calls and out-calls.

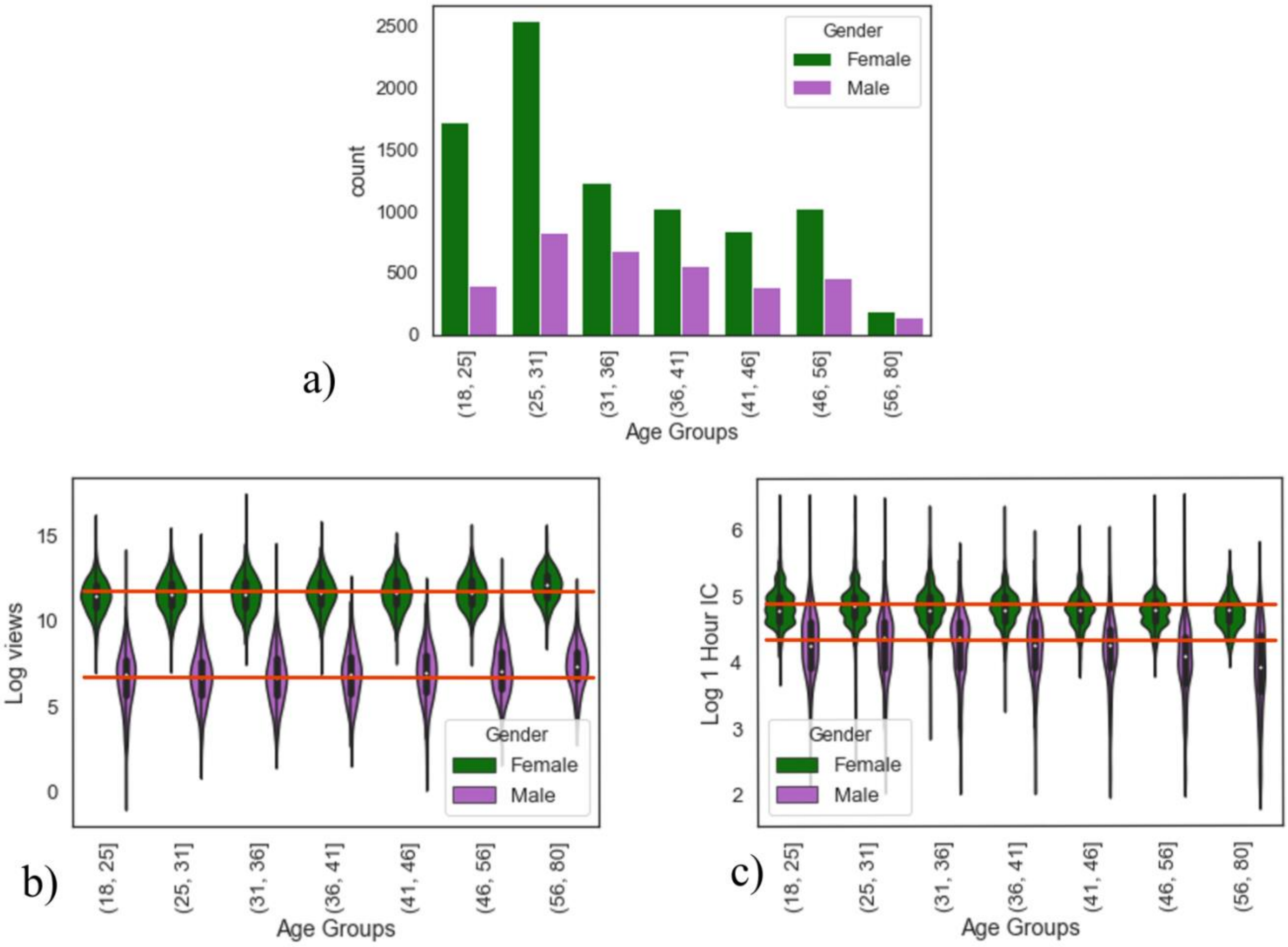

**Figure S3. Counts of workers by age group. b) views, and c) rates.**

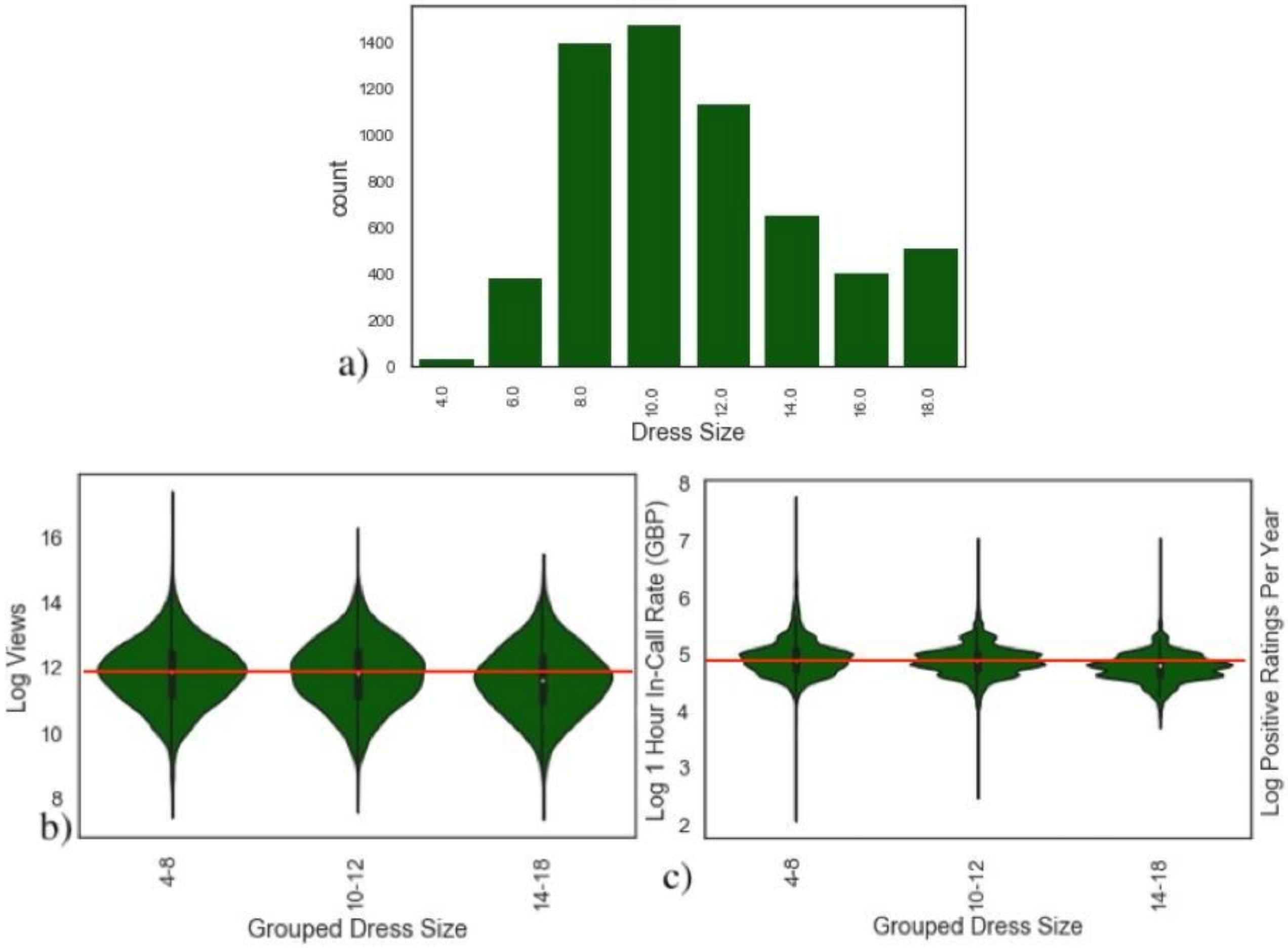

**Figure S4: a) Dress sizes count, b) views per day, and c) rates.**

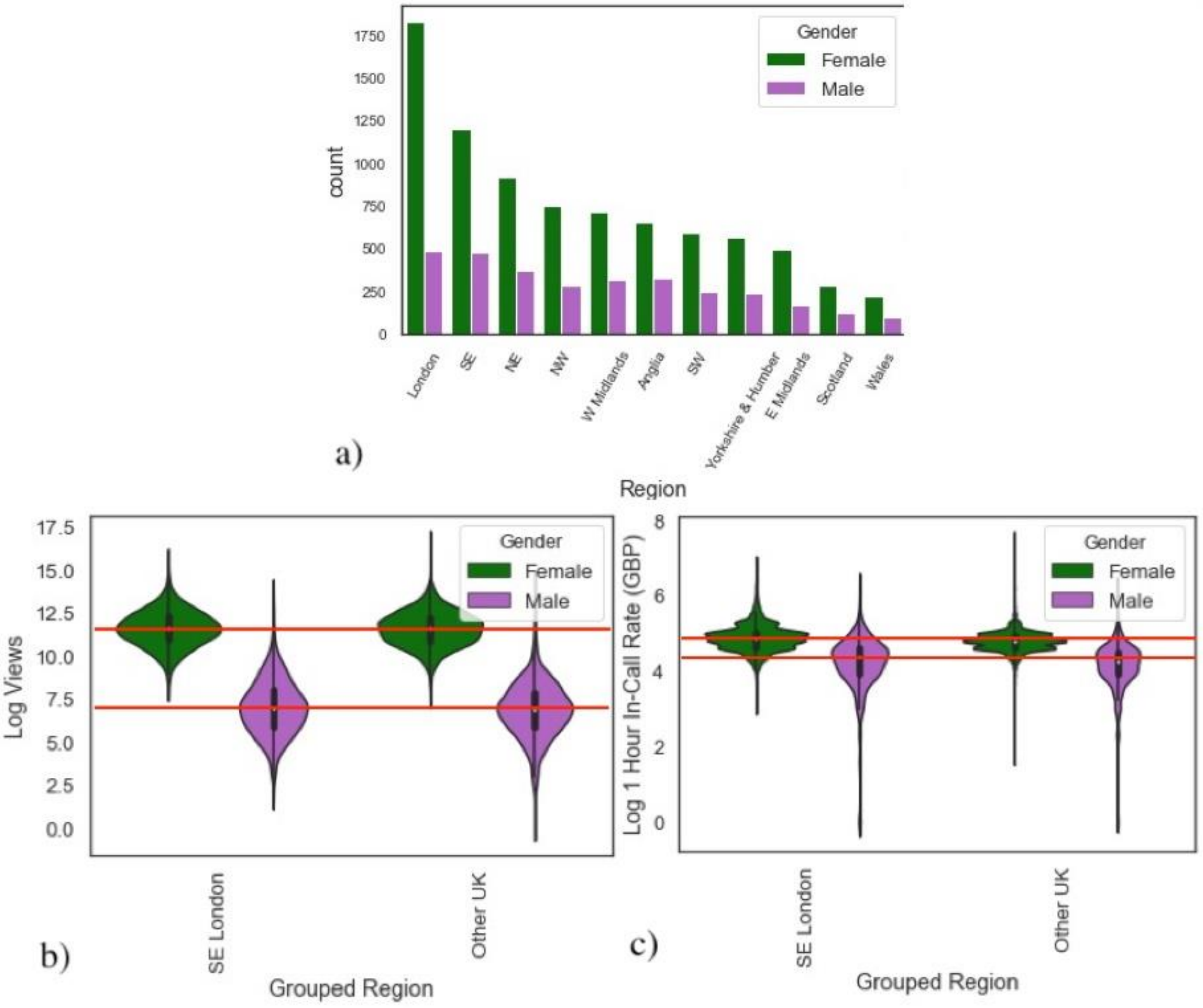

**Figure S5. a) Distribution of region, b) views, c) rates.**

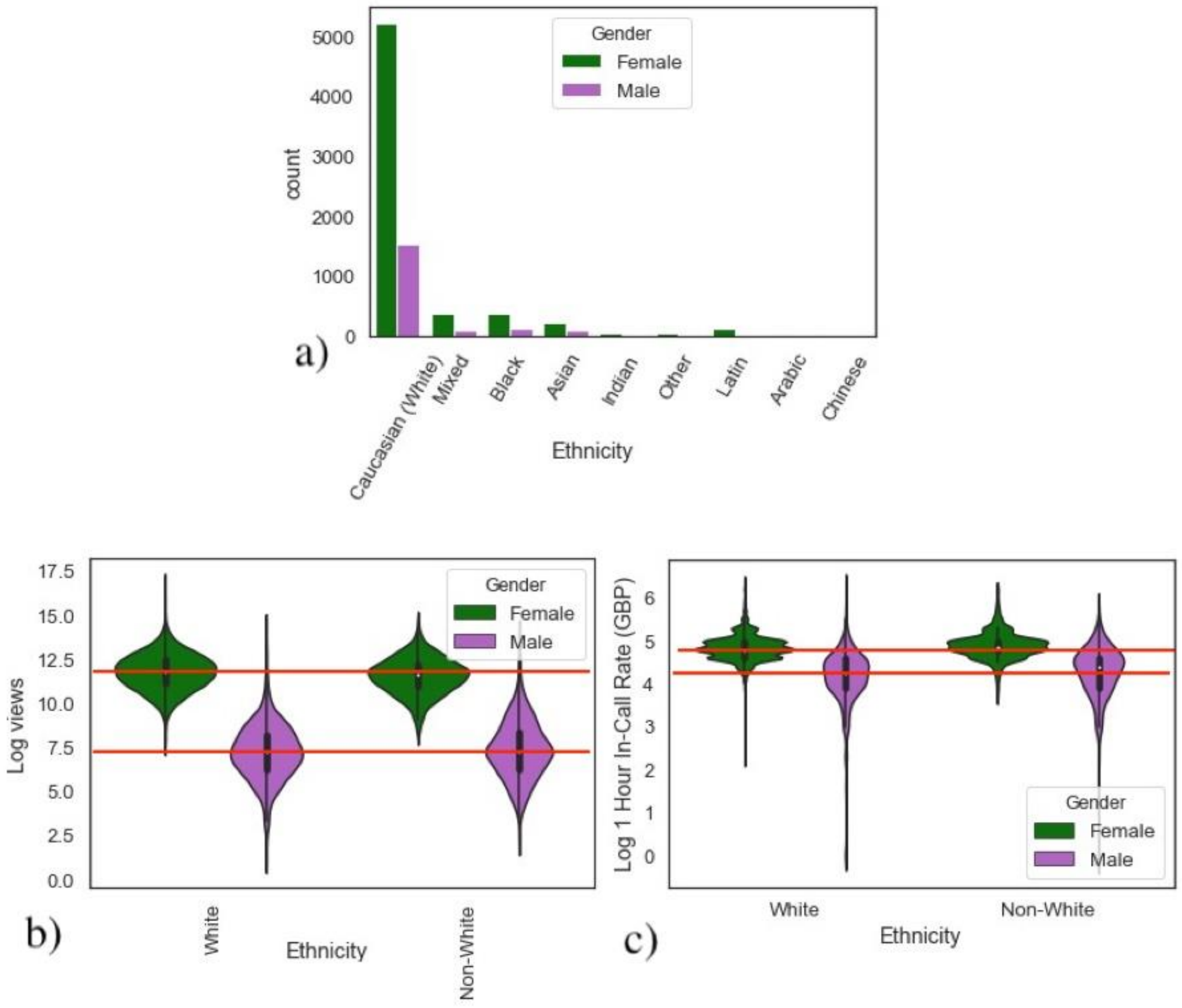

**Figure S6. a) Distribution of grouped ethnicity, b) views, c) rates.**

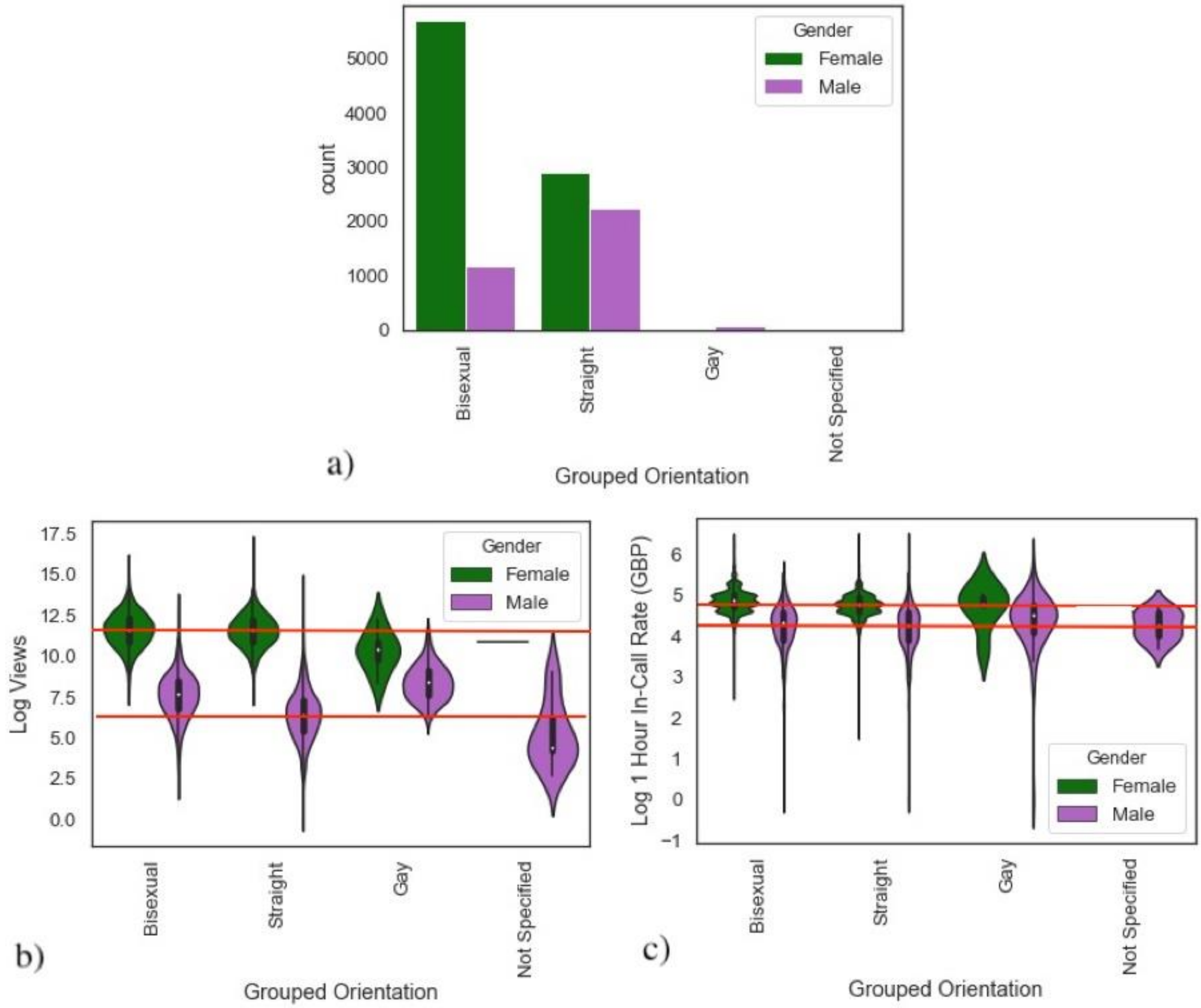

**Figure S7. a) Distribution of orientation b) views, c) rates.**

**Table S1. Grouped Nationalities**

| Grouped Nationality | Nationalities |
|---|---|
| Britain | British, British (Manx) |
| Middle East | Yemenite, Turkish, Iranian, Kuwaiti, ... |
| Eastern Europe | Romanian, Moldovan, Polish, Hungarian, Czech, Latvian, Lithuanian, Bulgarian, Slovakian, Russian, Slovenian, Ukrainian, Croatian, ... |
| Western Europe | Spanish, German, Maltese, Greek, Danish, Finnish, Swedish, French, Italian, Portuguese, Irish, Dutch, Icelander, Maltese, Austrian, Norwegian, ... |
| Asia | Chinese, Indian, Thai, Malaysian, Singaporean, Chinese (Hong Kong), Sri Lankan, Bangladeshi, Japanese, ... |
| Oceania | Filipino, Australian, New Zealander, ... |
| North America | American, Canadian |
| Central and South America | Mexican, Brazilian, Argentinian, Chilean, Colombian, Ecuadorian, ... |

**Table S2: Clothing measurement data ("Women's Clothing Size Conversion" 2017).**

| US & CAN | UK & AUS | EU | France | Bust | Waist | Hip |
|---|---|---|---|---|---|---|
| 4 | 8 | 34 | 36 | 32"/81 cm | 24"/61 cm | 35"/89 cm |
| 6 | 10 | 36 | 38 | 34"/86 cm | 26"/66 cm | 37"/94 cm |
| 8 | 12 | 38 | 40 | 36"/91 cm | 28"/71 cm | 39"/99 cm |
| 10 | 14 | 40 | 42 | 38"/97 cm | 30"/76 cm | 41"/104 cm |
| 12 | 16 | 42 | 44 | 40"/102 cm | 32"/81 cm | 43"/109 cm |
| 14 | 18 | 444 | 46 | 42"/107 cm | 34"/86 cm | 45"/114 cm |

## Table S3. Regression Results for Female Workers

| | | Views per day | Log 1 Hour In-call(£) |
|---|---|---|---|
| Age | | -0.208 (-1.47) | -0.00359*** (-10.26) |
| Ethnicity | Non-White | -13.54*** (-3.35) | -0.00254 (-0.26) |
| (baseline: White) | *Missing* | -32.49* (-2.01) | -0.0214 (-1.39) |
| Dress size | 10-12 | -15.35** (-3.09) | -0.0619*** (-6.99) |
| (baseline: 4-8) | 14-18 | -37.89*** (-6.39) | -0.156*** (-15.34) |
| | *Missing* | -14.63 (-0.84) | -0.0986*** (-6.08) |
| Nationality | Eastern Europe | 37.71*** (8.12) | -0.308*** (-31.60) |
| (baseline: British) | S./C. America | 32.35*** (3.81) | -0.0476* (-2.57) |
| | Asia | -2.450 (-0.47) | -0.229*** (-11.48) |
| | Western Europe | -1.320 (-0.21) | -0.0612* (-2.51) |
| | North America | -26.59** (-2.66) | 0.0758 (1.16) |
| | Africa | -10.63 (-1.02) | -0.0781* (-2.50) |
| | Oceania | -16.46 (-1.10) | 0.113* (2.45) |
| | Middle East | 11.89 (0.50) | 0.0547 (1.38) |
| | *Missing* | -19.18*** (-5.97) | -0.0841*** (-11.64) |
| Number of Services | | 3.107* (2.48) | -0.00901*** (-6.29) |
| Service Category | A | 8.271 (0.18) | 0.367*** (5.15) |
| | B | -122.7* (-2.21) | 0.452*** (6.64) |
| | C | -116.4 (-1.87) | 0.777*** (9.18) |
| | D | -82.93 (-1.38) | 0.460*** (5.97) |
| | E | 71.75 (0.85) | 0.119 (1.13) |
| Orientation | Bisexual | -0.631 (-0.22) | 0.0451*** (6.71) |
| (baseline: Straight) | Gay | -29.12 (-1.36) | 0.173 (0.50) |
| London & Southeast | | -0.952 (-0.32) | 0.107*** (15.95) |
| Views per day | | | |
| Log 1 Hour In-call(£) | | | |
| Constant | | 120.6*** (4.79) | 4.808*** (107.72) |
| Observations | | 8195 | 7990 |
| Adjusted $R^2$ | | 0.078 | 0.258 |

t-statistics in parentheses

* $p < 0.05$, ** $p < 0.01$, *** $p < 0.001$

## Table S4. Regression Results for Male Workers

| | | Views per day | Log 1 Hour In-call(£) |
|---|---|---|---|
| Age | | -0.0197 (-0.63) | -0.0175*** (-4.03) |
| Ethnicity | Non-white | 0.703 (0.62) | 0.0863 (0.86) |
| (baseline: White) | *Missing* | -1.017 (-1.83) | -0.155 (-1.20) |
| Nationality | Middle East | -3.665* (-2.24) | 0 (.) |
| (baseline: British) | North America | -3.187** (-2.86) | 0 (.) |
| | S./C. America | -4.396* (-2.50) | -0.795*** (-4.82) |
| | Asia | -4.686* (-2.37) | 0 (.) |
| | Africa | -1.466 (-0.59) | 0.281* (2.04) |
| | Eastern Europe | -2.556 (-1.26) | -0.205 (-1.89) |
| | Western Europe | -3.371 (-1.55) | 0.0934 (0.58) |
| | Oceania | 0 (.) | 0 (.) |
| | *Missing* | -4.042* (-2.24) | -0.260*** (-3.85) |
| Number of Services | | 0.0312 (0.43) | 0.0167 (1.79) |
| Service Category | A | -1.189 (-0.46) | -0.966* (-2.08) |
| | B | 2.479 (0.82) | -1.007 (-1.94) |
| | C | 3.891 (0.67) | -1.033 (-1.45) |
| | D | -5.083* (-2.08) | -0.662 (-1.26) |
| | E | 1.693 (0.31) | 0.562 (0.65) |
| Orientation | Bisexual | 0.832 (1.24) | 0.00645 (0.06) |
| (baseline: Straight) | Gay | 2.230 (1.80) | 0.211 (1.57) |
| London & Southeast | | -0.155 (-0.31) | 0.163* (2.16) |
| Views per day | | | |
| Log 1 Hour In-call(£) | | | |
| Constant | | 5.143** (2.63) | 5.917*** (12.34) |
| Observations | | 3115 | 2119 |
| Adjusted $R^2$ | | 0.011 | 0.150 |

t-statistics in parentheses

* $p < 0.05$, ** $p < 0.01$, *** $p < 0.001$

**Table S5. Logistic Regression for Unprotected Sex**

| | | Unprotected |
|---|---|---|
| Age | | -0.0177*** (-4.63) |
| Gender Male | | 1.537*** (18.85) |
| Ethnicity | Nonwhite | -0.398*** (-3.43) |
| (baseline: White) | *Missing* | 0.0937 (1.37) |
| Nationality | Eastern Europe | 0.307 ** (2.65) |
| (baseline: British) | S./C. America | -2.525* (-2.51) |
| | Western Europe | -0.891* (-2.44) |
| | Asia | -1.000* (-2.20) |
| | Africa | -0.864 (-1.21) |
| | Oceania | 0.00752 (0.01) |
| | North America | 0 (.) |
| | Middle East | 0 (.) |
| | *Missing* | 0.0227 (0.28) |
| Orientation | Bisexual | 0.158* (2.35) |
| (baseline: straight) | Gay | 0.245 (0.81) |
| London & Southeast | | -0.00244 (-0.04) |
| TV-TS | | 0.445 (1.67) |
| Constant | | -2.196*** (-13.81) |
| Observations | | 11489 |

t statistics in parentheses
* $p < 0.05$, ** $p < 0.01$, *** $p < 0.001$